\documentclass[prl,twocolumn,superscriptaddress,tightenlines,showpacs]{revtex4}
\usepackage{amssymb}
\usepackage{epsfig}
\usepackage{amsbsy}
\usepackage{amsmath}
\usepackage{graphicx}
\usepackage{CJK}
\usepackage{amsfonts}

\begin{document}

\title{Sensitive Chemical Compass Assisted by Quantum Criticality }

\author{C. Y. Cai}
\affiliation{Institute of Theoretical Physics, Chinese Academy of Sciences, Beijing, 100190, China}

\author{Qing Ai}
\affiliation{Institute of Theoretical Physics, Chinese Academy of Sciences, Beijing, 100190, China}

\author{H. T. Quan}
\affiliation{Department of Chemistry and Biochemistry and Institute for Physical Science
and Technology, University of Maryland, College Park, MD 20742, USA}

\author{C. P. Sun}
\affiliation{Institute of Theoretical Physics, Chinese Academy of Sciences, Beijing, 100190, China}
\email{suncp@itp.ac.cn}
\homepage{http://power.itp.ac.cn/ suncp/index.html}

\begin{abstract}
The radical-pair-based chemical reaction could be used by birds for the
navigation via the geomagnetic direction. An inherent physical mechanism is
that the quantum coherent transition from a singlet state to triplet states
of the radical pair could response to the weak magnetic field and be
sensitive to the direction of such a field and then results in different
photopigments in the avian eyes to be sensed. Here, we propose a quantum
bionic setup for the ultra-sensitive probe of a weak magnetic field based on
the quantum phase transition of the environments of the two electrons in the
radical pair. We prove that the yield of the chemical products via the
recombination from the singlet state is determined by the Loschmidt echo of
the environments with interacting nuclear spins. Thus quantum criticality of
environments could enhance the sensitivity of the detection of the weak
magnetic field.
\end{abstract}

\maketitle


%
%
%


\textit{Introduction.---} Since Schr\"{o}dinger questioned \textquotedblleft
what is life" from the general point view of a quantum physicist~\cite%
{schrodinger}, scientists have never stopped the long-term exploration for
the physical sources of the living phenomena, and this even stimulated the
enthusiasm for the great discovery of the DNA genetic molecule~\cite{derry}.
Today it seems trivial to say that the life is of quantum for the molecules
composing lives obey quantum laws, but some recent discoveries are very
intriguing as some optimized living processes may be based on the $\mathit{%
nontrivial}$ quantum effect from quantum coherence. One example in point is
the photosynthesis process. Recent experiments have been able to exactly
determine the time scales of various transfer processes by the 2D optical
spectroscopy, and then show quantum coherence effects in energy transfer via
collective excitations of some light-harvesting complexes~\cite%
{fleming1,cao,yang-xu-sun}.

Another prototype of quantum coherence effect for living process seems to
appear in the avian magnetoreception mechanisms~\cite{johnsen,ritz1}
verified by some recent experiments~\cite{qcoherenceexp}. Recently quantum
information approaches have been used to further analyze the role of quantum
coherence phenomenon in the avian magnetoreception models~\cite%
{kominis,cai1,vedral,cai2}. It is now believed~\cite{pnas09,pureapplchem09} that the
model for magnetoreception is based on the radical-pair mechanism (RPM)~\cite%
{ritz1}: the radical-pair molecule with two unpaired electrons is activated
by light. When the electrons interact with their individual environments of
nuclei via the hyperfine couplings, the spin singlet state will transit to
the spin triplet states even though the external field is uniform and rather
weak. In response to this quantum coherent transition, the field-dependent
change in the product yield of the radical-pair-based chemical reaction is
enough to be sensed by the avian retina.

\begin{figure}[tbh]
\includegraphics[width=6cm, clip]{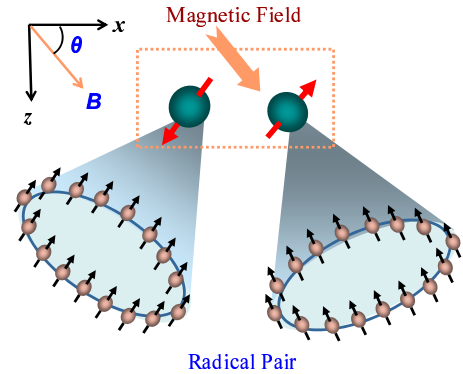}
\caption{ (color online). Model setup of sensitive magneto-detection based
on the radical pair chemical reaction assisted by the quantum critical
environments consisting of interacting spins. Two spins could be initially
prepared in the singlet state, and finally evolve into the triplet states
due to the couplings to their environments. The corresponding chemical
product sensitively responses to the external magnetic field.}
\label{scheme}
\end{figure}

In spite of the rapid progress in the understanding of the RPM in the last
decade, a key question remains elusive~\cite{pnas09}: why does the
singlet-triplet interconversion response to the extremely-weak geomagnetic
field ($\sim 10^{-5}$\textrm{T}), and why is it very sensitive to its
direction? We notice that the existence of nuclear environments surrounding
the electron spins in the radical-pair molecule is crucial to magnetic
sensitivity of the chemical reaction. This observation motivates us to
consider the role of internal quantum correlation in each environment. In
this Letter, we propose a quantum phase transition (QPT)-assisted setup for
the probe of a weak magnetic field in understanding the above conundrum.
Actually a lot of real-world detectors are built based on the dramatic
changes of the systems around phase transitions, which amplifies the
ultra-weak signal and thus enable one to probe it. Examples include bubble
chamber detectors~\cite{bubble} and superconducting single-photon detectors~%
\cite{singlephoton}, where the liquid-gas phase transition and
superconductor-metal phase transition take place respectively enhancing the
sensitivity for detection. We calculate the corresponding chemical product
yield, which is phenomenologically described by a damping process~\cite%
{steiner89}. It is discovered that the chemical product yield is determined
by the time integral of the Loschmidt echo (LE). Our result shows that the
chemical compass assisted by quantum criticality indeed can response to a
very-weak magnetic field and be sensitive to its direction.

\textit{Quantum-Phase-Transition-Assisted Radical Pair Mechanism.---} Our
setup is illustrated in Fig.~\ref{scheme}. Each of the two electrons in the
radical pair is uniformly coupled to its own quantum correlated environment,
which can be described by a transverse field Ising (TFI) model~\cite{qpt,lmg}%
. In an external magnetic field $B(\cos \theta \hat{x}+\sin \theta \hat{z})$%
, which may be the geomagnetic field, the environment is described by $%
H_{n}^{\prime }=H_{n}+g_{N}\mu _{N}B\sum\nolimits_{j}\sin \theta I_{n,j}^{z}$
with
\begin{equation}
H_{n}=J\sum\limits_{j=1}^{N}(I_{n,j}^{z}I_{n,j+1}^{z}+\lambda I_{n,j}^{x}),
\end{equation}%
where $\lambda =g_{N}\mu _{N}B\cos \theta /J$ is the rescaled strength of
the transverse field in unit of $J$ being the Ising coupling constant, $%
g_{N}\mu _{N}$ is the nuclear magnetic moment, and $n=1,2$ refers to the
environment of the $n$th electron. $I_{n,j}^{x}$ and $I_{n,j}^{z}$ are Pauli
matrices of the $j$th nuclear spin operators. In case of antiferromagnetic
Ising chain, i.e., $J>0$, we can omit the longitudinal terms and $%
H_{n}^{\prime }\approx H_{n}$ since they lead to only higher-order
correction~\cite{qptexp2}. Here, different from all previous studies~\cite%
{ritz1,pnas09,vedral,pureapplchem09,steiner89,Hor07}, we explicitly consider
the inter-nucleus couplings $I_{n,j}^{z}I_{n,j+1}^{z}$. This coupling
competes with the Zeeman energy being proportional to the geomagnetic field,
and leads to a QPT at the critical point $\lambda =1$. A central spin
uniformly coupled with each spin in this TFI system possesses the dynamic
sensitivity described by the sharp decay of LE near the QPT~\cite{quan-sun},
which has been experimentally verified~\cite{qptexp1,qptexp2,qptexp3}.

The two unpaired electrons in the radical pair couple to the two
environments $E_{1}$ and $E_{2}$ respectively with the following
Hamiltonians
\begin{equation}
V_{n}\!=\!\Omega \sin \theta \sigma _{n}^{z}+\Omega \cos \theta \sigma
_{n}^{x}+Jg\sigma _{n}^{x}\sum\nolimits_{j}I_{n,j}^{x},
\end{equation}%
where $\sigma _{n}^{x}$ and $\sigma _{n}^{z}$ for $n=1,2$ are the Pauli
operators for the $n$th electron spin, the dimensionless coupling constant
scales as $g=g_{0}/\sqrt{N}$ in the van Hove limit for the interacting
many-body system. All the information about the geomagnetic field is also
incorporated in $\theta $ and $\Omega $, which is the electronic Zeeman
energy splitting induced by the geomagnetic field.

\textit{Radical-Pair Evolution in Correlated Nuclear Environment.---} Due to
the spin-flip terms, the time evolution governed by the total Hamiltonian $%
H=\sum_{n}(H_{n}+V_{n})$ can only be solved with some approximation. Usually
the electron spins evolve faster than the nuclear spins. Thus we can first
regard nuclear spins as c-numbers for formally diagonalizing the electronic
Hamiltonian $V_{n}$ through a generalized Born-Oppenheimer approximation~%
\cite{sun00}. The eigen states of the electron spins are obtained as $%
\left\vert +\right\rangle =\cos (\alpha /2)\left\vert \uparrow \right\rangle
+\sin (\alpha /2)\left\vert \downarrow \right\rangle $ and $\left\vert
-\right\rangle =\sin (\alpha /2)\left\vert \uparrow \right\rangle -\cos
(\alpha /2)\left\vert \downarrow \right\rangle $ where $\left\vert \uparrow
\right\rangle $ ($\left\vert \downarrow \right\rangle $) is the spin-up
(down) state in the $\sigma _{x}$-representation with corresponding eigen
values $E_{\pm }=\pm E$ for $E=\sqrt{\Omega ^{2}\sin ^{2}\theta +\Delta ^{2}}
$ and $\Delta =\Omega \cos \theta +Jg\sum\nolimits_{j}I_{n,j}^{x}.$ And the
mixing angle is defined as $\alpha =\pi /2-\tan ^{-1}\left( \Omega \sin
\theta /\Delta \right) .$

For the weak coupling $(Jg_{0}\ll \Omega )$ of an electron to nuclei we
shall approximately obtain the eigen states $\left\vert +\right\rangle
\simeq \cos \theta ^{\prime }\left\vert \uparrow \right\rangle +\sin \theta
^{\prime }\left\vert \downarrow \right\rangle $ and $\left\vert
-\right\rangle \simeq \sin \theta ^{\prime }\left\vert \uparrow
\right\rangle -\cos \theta ^{\prime }\left\vert \downarrow \right\rangle $
to the zeroth order of $g$ for $\theta ^{\prime }=\left( \pi /2-\theta
\right) /2$ and the eigen energy $E\simeq \Omega +Jg\cos \theta
\sum\nolimits_{j}I_{n,j}^{x}$ to the first order. The Born-Oppenheimer
approximation shows that the slowly-varying nuclear spins would not induce
the coherent transition of the fast-varying electronic degrees, but the
electronic motion provides an effective potential for the nuclear spins. In
this sense, the total Hamiltonian is approximately rewritten as
\begin{equation}
H\simeq \sum\nolimits_{n}\left( H_{n}^{+}\left\vert +\right\rangle
\left\langle +\right\vert +H_{n}^{-}\left\vert -\right\rangle \left\langle
-\right\vert \right) ,
\end{equation}%
where the different effective Hamiltonians for the nuclear spins
corresponding to states $\left\vert \pm \right\rangle $\ are respectively%
\begin{equation}
H_{n}^{\pm }=J\sum\nolimits_{j}\left[ I_{n,j}^{z}I_{n,j+1}^{z}+(\lambda \pm
g\cos \theta )I_{n,j}^{x}\right] \pm \Omega .
\end{equation}
It can be proven that the Born-Oppenheimer approximation is generally valid
even for such a large $N$ that a QPT occurs.

\textit{Product Yield and Loschmidt Echo.---} The radical pair is assumed to
be initially in the singlet state $\left\vert S\right\rangle $ which
subsequently suffers from the homogeneous interaction $V=\sum_{n}V_{n}$ with
the environmental nuclear spins. Then the radical pair undergoes a
singlet-to-triplet transition. The charge recombination of the radical pair
goes through different channels, depending on the electron-spin state
(singlet or triplet). In particular, the singlet-state product yield formed
by the reaction of radical pairs can be calculated as~\cite{steiner89}
\begin{equation}
\Phi _{S}(t)=\int_{0}^{t}r_{c}(t)f_{S}(t)dt,
\end{equation}%
where $r_{c}(t)$ is the radical re-encounter probability distribution, and $%
f_{S}(t)=\langle S\left\vert \rho _{e}(t)\right\vert S\rangle $ is the
singlet-state population at time $t$. Usually it is assumed \cite{steiner89}
that $r_{c}(t)=k_{S}\exp {(-k_{S}t)}$ with $k_{S}$ the recombination rate.
The ultimate product yield $\Phi _{S}\equiv \Phi _{S}(t\rightarrow \infty )$
in cryptochrome is believed to affect the visual function of animals~\cite%
{ritz1}. In order to quantitatively describe the magnetic-field sensitivity
of the radical-pair reaction, we shall resort to $\Lambda (\theta )=\partial
\Phi _{S}/\partial \theta $,$\,$the derivative of the product yield with
respect to the geomagnetic-field direction $\theta $~\cite{Hor07}.

For nuclear spins initially in the mixed state $\rho _{1}\otimes \rho _{2}$,
the initial state of the total system is $\rho (0)=\left\vert S\right\rangle
\left\langle S\right\vert \otimes \rho _{1}\otimes \rho _{2}$. When the
state of the electron is $\left\vert +\right\rangle $ or $\left\vert
-\right\rangle $, the evolution of the environment is governed by $H_{n}^{+}$
or $H_{n}^{-}$ respectively, and hence the initial state $\rho _{n}$ will
evolve into $U_{n}^{+}\rho _{n}\left( U_{n}^{+}\right) ^{\dag }$ or $%
U_{n}^{-}\rho _{n}\left( U_{n}^{-}\right) ^{\dag }$ respectively. Here, the
evolution operators are defined as $U_{n}^{\pm }=\exp \left( -iH_{n}^{\pm
}t\right) $. This will result in the so-called adiabatic entanglement
between the system and the environment \cite{sun00}. For the above initial
state $\rho (0)$, at time $t$, the reduced density matrix for the electron
spins $\rho _{e}(t)=$tr$_{n}\left\vert \psi (t)\right\rangle \left\langle
\psi (t)\right\vert $ reads
\begin{eqnarray}
\rho _{e}(t) &=&\frac{1}{2}[\left\vert +-\right\rangle \left\langle
+-\right\vert +\left\vert -+\right\rangle \left\langle -+\right\vert  \notag
\\
&&-D(t)\left\vert +-\right\rangle \left\langle -+\right\vert -D^{\ast
}(t)\left\vert -+\right\rangle \left\langle +-\right\vert ],
\end{eqnarray}%
where $D(t)=\mathrm{tr}\left[ U_{1}^{+}\rho _{1}\left( U_{1}^{-}\right)
^{\dag }\right] \mathrm{tr}\left[ U_{2}^{-}\rho _{2}\left( U_{2}^{+}\right)
^{\dag }\right] $ is the decoherence factor. When $\rho _{1}$ $=\rho
_{2}=\rho $, $D(t)$ is real and can be simplified to $L(t)=\left\vert
\mathrm{tr}\left[ U^{+}\rho \left( U^{-}\right) ^{\dag }\right] \right\vert
^{2}$, which is just the \textit{Loschmidt Echo }characterizing the dynamic
sensitivity of the environment in response to the perturbation \cite%
{quan-sun}. Straightforwardly, we can prove the population $f_{S}(t)=\left[
1+L(t)\right] /2$, and obtain the product yield as
\begin{equation}
\Phi _{S}=\frac{1}{2}+\frac{1}{2}k_{S}\int\nolimits_{0}^{\infty
}L(t)e^{-k_{S}t}dt.  \label{PhiS}
\end{equation}%
This is a central result of our paper which reveals the direct relationship
between the product yield and the LE.

Hereafter, we calculate the product yield for identical environments in an
initial pure state, which corresponds to the case with absolute zero
temperature. In this case, the initial state can be described by a state
vector $\left\vert G\right\rangle $ and LE is simplified as $L(t)=\left\vert
\left\langle G\right\vert \exp \left( iH^{-}t\right) \exp \left(
-iH^{+}t\right) \left\vert G\right\rangle \right\vert ^{2}$, which was
explicitly given in Ref.~\cite{quan-sun}. In the following we take two
specific cases into account. First, 
we consider the large-$N$ case since the QPT usually occurs in this limit.
Using the analytic results about $L(t)$ obtained in Ref.~\cite{quan-sun}, we
approximate the product yield around the critical point as
\begin{equation}
\Phi _{S}\simeq \frac{1}{2}+\exp \left( \frac{k_{S}^{2}}{4\gamma }\right)
\sqrt{\frac{\pi k_{S}^{2}}{2\gamma }}\left[ 1-\text{erf}\left( \frac{k_{S}}{2%
\sqrt{\gamma }}\right) \right] ,
\end{equation}%
where $\mathrm{erf}(x)\ $is the error function, and
\begin{equation}
\gamma =\frac{8J^{2}g^{2}NK_{c}^{3}\cos ^{2}\theta }{3\pi (1-\lambda )^{2}}
\end{equation}%
with $K_{c}$ the cutoff momentum. For a sufficiently small $k_{S}$, i.e., $%
k_{S}\ll 2\sqrt{\gamma }$, the product yield is approximated as
\begin{equation}
\Phi _{S}\approx \frac{1}{2}+\frac{\pi k_{S}\left\vert 1-\lambda \right\vert
}{16Jg\cos \theta }\sqrt{\frac{6}{NK_{c}^{3}}}.
\end{equation}%
Since $\partial \Phi _{S}/\partial \lambda $ is discontinuous when $\lambda
=1$, it may serve as a witness of the QPT.

To the opposite, we consider the case with a small $N$, in which the above
analysis fails. Therefore, we shall deal with it separately. For $N=2$, we
obtain explicitly
\begin{equation}
L(t)=1-\frac{16g^{2}\cos ^{2}\theta \sin ^{2}(\xi t)}{\left[ 1+4(\lambda
-g\cos \theta )^{2}\right] \xi ^{2}},
\end{equation}%
where $\xi =\sqrt{1+4(\lambda +g\cos \theta )^{2}}$. It shows that the setup
for a small $N$ can not work as well as it for a large $N$. The detailed
discussions about dependence of $\Phi _{S}$ on $N$ are visually given in the
supplementary material.
\begin{figure}[tbh]
\includegraphics[width=8cm, clip]{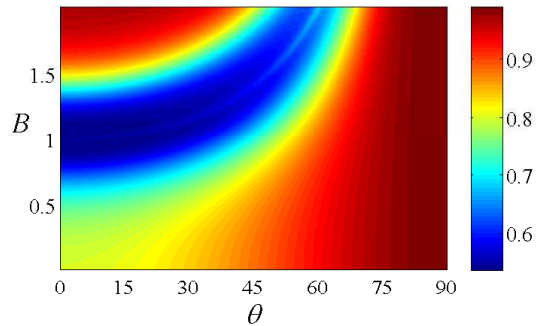}
\caption{ (color online). The product yield $\Phi _{S}$ vs the magnitude $B$
and direction $\protect\theta $ of the magnetic field at a finite
temperature $T=0.2$ with $N=1000$, $g_{0}=1$, $k_{S}=0.1$, and $J=1$.}
\label{PhiSvsThetaLambda}
\end{figure}

\textit{Sensitive Magnetodetection at Finite Temperatures.---} The above
results are obtained for an ideal case with pure states. For the practical
purpose, we need to consider the cases at finite temperatures. In order to
illustrate the result for a very-large $N$, we numerically plot the product
yield $\Phi _{S}$ vs the magnitude $B$ and direction $\theta $ of magnetic
field at a finite temperature in Fig.~\ref{PhiSvsThetaLambda}. Obviously,
the product yield displays its dependence on both the geomagnetic field's
magnitude and direction. Besides, there is a deep valley around the top-left
corner. This can be seen from the fact that the LE decays in a gaussian way
around the critical point $\lambda \simeq 1$. Additionally, in the regions
far away from the critical point, e.g., at the top-left and bottom-right
corners, the product yield nearly stays unity for the LE almost does not
decay.

\begin{figure}[tbh]
\includegraphics[width=9cm, clip]{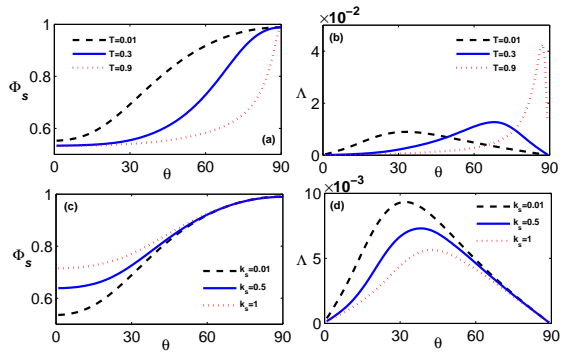}
\caption{ (color online). The product yield $\Phi _{S}$ and its derivative $%
\Lambda$ vs the geomagnetic field's direction $\protect\theta $ for
different $T$ in (a) and (b) with $k_{S}=0.1$, and for different $k_{S}$ in
(c) and (d) with $T=0.01$. In all figures, we set $N=1000$, $g_{0}=1$, $%
B=0.9 $ and $J=1$. }
\label{PhiSvsFixed3Para}
\end{figure}

Furthermore, in order to investigate the influence of other parameters, we
plot the product yield vs direction for different temperatures and
recombination rates in Fig.~\ref{PhiSvsFixed3Para}. The similarity among the
cases with different temperatures is that there is a peak for $\Lambda
(\theta )$ as it increases from zero at $\theta =0$. It is seen that as the
temperature increases, the position of the peak moves towards $\theta =\pi
/2 $, meanwhile the line shape on the left hand side becomes more and more
flat. In the high-temperature limit, we would expect a sharp peak around $%
\theta =\pi /2$, while there is a platform elsewhere. In this case, the bird
can no longer discriminate the direction. This is a reasonable result since
a QPT takes place at the absolute zero, and a high temperature smears the
QPT. When we come to the recombination rate $k_{S}$ in Fig.~\ref%
{PhiSvsFixed3Para}(c), we see that the slower the singlet state reacts, the
more the product yield changes along with the direction. That is because a
reaction with a smaller recombination rate provides more time for the decay
of LE. The magnetic-field sensitivity in Fig.~\ref{PhiSvsFixed3Para}(d)
clearly confirms our analysis. In addition, as the environment involves more
nuclear spins, the visibility rises as the LE decays faster for a larger $N$.

\textit{Conclusion.---} We proposed a RPM-based magnetodetection scheme
assisted by the QPT. We have proved that the yield of the chemical product
via the recombination from the singlet state is determined by the LE of the
environments. This relation results in the enhanced sensitivity of the
RPM-based avian compass. Thus, our study does not only provide important
insights to the mechanism of magnetoreceptor through a radical-pair process
in a very-weak field, but it also shields new light on building a quantum
bionic device for ultra-sensitive magnetic-field sensing. In addition, in
our bionic setup, the sensitivity is pronounced when the nuclear spin number
is large and the recombination rate is small.

It may be argued that in the avian retina, the environments of radical pairs
consist of a few nucleus rather than numerous nucleus, i.e., $N\rightarrow
\infty $ for a QPT. But the experiments~\cite{qptexp1,qptexp2,qptexp3}
however demonstrated that even for $N=2$, there still exists the dynamic
sensitivity induced by quantum criticality. This result implies that dynamic
sensitivity may have a close relation with the level crossing. Besides,
although our scheme requires a very-low temperature for the current
experimental parameters, its importance also lies in the possible bionic
setup for sensitive magnetodetection. Last but not least, our results are
based on the TFI model, but the enhancement of LE decay due to QPT is
independent of the model~\cite{fernando07}. Thus it is reasonable to infer
that all the results obtained from TFI model can be generalized to other
models as well, such as the so-called long-range TFI model, or
Lipkin-Meshkov-Glick model~\cite{lmg}. Detailed studies of these
generalization will be given in a forthcoming paper.

The work is partially supported by National Natural Science Foundation of
China under Grant Nos. 10935010 and 11074261. H.T.Q. is supported by the
National Science Foundation (USA), grant DMR-0906601.

\end{document}